\documentclass[a4paper,10pt]{article}
\usepackage{enumerate}
\usepackage{color}
\usepackage[utf8]{inputenc} 
\usepackage[english]{babel}
\usepackage[T1]{fontenc}
\usepackage{graphicx}
\usepackage{amsfonts,amssymb,amsmath,latexsym,amsthm}
\usepackage{textcomp}
\usepackage[pdftex]{hyperref}
\usepackage{geometry}
\title{Soliton dynamics in random  fields: The Benjamin-Ono equation framework}
\author{Marcelo V. Flamarion$^{1}$, Efim Pelinovsky$^{2,3}$ and Ekaterina Didenkulova$^{2,3}$}
\date{}

\begin{document}
\maketitle
\begin{center}
{\footnotesize $^1$Departamento Ciencias—Secci{\' o}n Matem{\' a}ticas, Pontificia Universidad Cat{\' o}lica del Per{\' u}, Av. Universitaria 1801, San Miguel 15088, Lima, Peru  \\
mvellosoflamarionvasconcellos@pucp.edu.pe}

\vspace{0.3cm}
{\footnotesize 
$^{2}$Faculty of Informatics, Mathematics and Computer Science, HSE University, Nizhny Novgorod 603155, Russia.

$^{3}$ Gaponov-Grekhov Institute of Applied Physics, Nizhny Novgorod, 603122, Russia.}




\end{center}


\begin{abstract} 

Algebraic soliton interactions with a periodic or quasi-periodic random force are investigated using the Benjamin-Ono equation. The random force is modeled as a Fourier series with a finite number of modes and random phases uniformly distributed, while its frequency spectrum has a Gaussian shape centered at a peak frequency. The expected value of the averaged soliton wave field is computed asymptotically and compared with numerical results, showing strong agreement. We identify parameter regimes where the averaged soliton field splits into two steady pulses and a regime where the soliton field splits into two solitons traveling in opposite directions. In the latter case, the averaged soliton speeds are variable. In both scenarios, the soliton field is damped by the external force. Additionally, we identify a regime where the averaged soliton exhibits the following behavior: it  splits into two distinct solitons and then recombines to form a single soliton. This motion is periodic over time.
	\end{abstract}

\section{Introduction}

Solitons are coherent, localized waves with particle-like behavior that travel long distances while maintaining their shape and speed. They occur in problems when there is a perfect balance between dispersion and nonlinearity. Solitons have applications in various fields of research, including water waves \cite{Mignerey:2003, Duda:2004, Apel:2007}, plasma physics, biology, quantum physics, and more \cite{Joseph:2016}. The term "soliton" was coined in the seminal work of Zabusky and Kruskal \cite{Zabusky:1965}. While studying solitary wave collisions for the Korteweg-de Vries (KdV) equation, researchers observed that after interacting, the two waves regained their initial shape and speed. However, a phase lag occurred, indicating that the crests of the waves were shifted from the trajectories of their incoming centers. Rigorous results on soliton interactions for the KdV equation were later proved by Lax \cite{Lax:1968}. Since then, many researchers have investigated soliton interactions using different models. Examples include solitons in internal waves in shallow water regimes \cite{Grimshaw:2010a, Grimshaw:2010, Bokaeeyan:2019}, solitons in the Intermediate Long Wave (ILW) equation for fluids of finite depth \cite{Choi:1996, Matsuno:1993a, Matsuno:1993b}, and solitons in non-integrable frameworks such as the Euler equations \cite{Craig:2006}, the Whitham equation \cite{Flamarion:2022}, and the Schamel equation \cite{Flamarion-Pelinovsky-Didenkulova:2023}. The concept of generalized solitons, which have tails that decay at infinity \cite{Malomed:1996}, has also appeared in the literature, particularly in the context of gravity-capillary flows. Algebraic solitons, such as those described by the Benjamin-Ono (BO) equation \cite{Benjamin:1967, Davis:1967, Ono:1975, Ko:1978, Choi:1996, Meng:2014, Bona:2020, Bona:2023, Matsuno:1984, Yoneyama:1986}, are another notable example. Understanding soliton interactions paves the way for the study of soliton gas theory, which initially emerged within integrable models such as the nonlinear Schr{\" o}dinger and KdV equations. The terms "soliton gas" or "soliton turbulence" typically refer to an ensemble of solitons with random parameters within these integrable systems \cite{Zakharov:1971, Zakharov:2009}, however it has also been considered in non-integrable systems recently \cite{Flamarion-Pelinovsky-Didenkulova:2023b}.

In many physical problems, external forces can significantly influence the water wave dynamics. For instance, variable atmospheric pressure and topographic obstacles can affect these dynamics \cite{Ermakov, Flamarion:2022, Grimshaw86, Grimshaw:1993, Malomed:1993, Pelinovsky:2002, Smyth:2002, Wu1, Flamarion:2021, COAM2, Flamarion:2022b, Flamarion:2022c, Flamarion:2022d}. When the external force depends on space, a resonance between the external force and the wave field can arise, potentially generating trapped waves \cite{Grimshaw94, Grimshaw96}. However, in many cases, the external force cannot be considered deterministic \cite{Debussche:1999, Liu:2007, Pelinovsky:2006, Wadati:1983, Wadati:1984, Wadati:1990, Zahibo:2009}. In this context, Zahibo et al. \cite{Zahibo:2009} considered the weakly damped KdV equation under the influence of a random time-dependent external force and showed that the averaged soliton field over long periods transforms into a "thick" soliton or a KdV-like soliton, depending on the statistical properties of the force. Sergeeva and Pelinovsky \cite{Pelinovsky:2006} examined the KdV solitons influenced by a time-dependent periodic external force. They found that the amplitude of the averaged soliton field decays with increasing frequency, and in certain regimes, the averaged soliton splits into two pulses propagating at the same velocity but in different directions. While the effects of a random force has been discussed in the KdV framework, as far as we know  there are no studies considering the influence of a random periodic field on algebraic solitons for BO equation.

In this work, we investigate the BO equation, which is described by
\begin{equation}\label{BOR}
\eta_{t} + \eta\eta_{x} + \mathcal{H}[\eta_{xx}] = F(t),
\end{equation}
where $F(t)$ is in general a quasi-periodic force with random phases.

The BO equation frequently appears in the literature as a model for the interface between two inviscid fluids of constant densities \cite{Benjamin:1967, Bona:2020, Bona:2023, Choi:1996, Davis:1967, Flamarion-Pelinovsky:2023, Matsuno:1982a, Matsuno:1982b, Matsuno:1993a, Matsuno:1993b, Ono:1975}. This interface is characterized by a flat, rigid lid and infinite depth. Here, the $\eta(x,t)$ represents undisturbed interface elevation at position $x$ and time $t$. The term $\mathcal{H}$ denotes the Hilbert transform defined as
\begin{equation}\label{Hilbert}
\mathcal{H}[\eta(x,t)]=\frac{1}{\pi}\int_{-\infty}^{+\infty}\frac{\eta(z,t)}{z-x}dz.
\end{equation}
The random external force $F(t)$, which can be associated with disturbances in the interface \cite{Choi:1996} is modeled as a Fourier series with a finite number of modes and random phases uniformly distributed within a bounded interval. Its frequency spectrum follows a Gaussian shape centered at a specific peak frequency.  

In this work, we utilize asymptotic theory to compute the averaged soliton field and validate our findings through comparison with numerical simulations, achieving strong agreement. Our analysis uncovers regimes where the soliton field undergoes a notable splitting phenomenon. In one scenario, the split solitons maintain steadiness, exhibiting an amplitude inversely proportional to the force frequency. Conversely, in another scenario, the soliton amplitude decays over time, while the split solitons propagate in opposite directions at varying speeds. Additionally, we identify a regime in which the soliton periodically splits into two and then rejoins to form a single soliton over time. To the best of our knowledge these phenomena are reported here for the first time in the literature.

For reference, this article is organized as follows: In Section 2 we reformulate the problem into a homogenous equation. In Section 3, we describe the main results on the characteristics of the averaged soliton field and the soliton dynamics in the presence of a random periodic field. Then, we present the final considerations in Section 4.

%
%

\section{Mathematical modeling}

We transform equation (\ref{BOR}) into an unforced problem with constant coefficients, we utilize the quadratic nonlinearity of the BO equation, following the methodology presented by Zahibo et al. \cite{Zahibo:2009}. We start with the change of variables:
\begin{equation}\label{Ch1}
\eta(x,t)=\zeta(x,t)+Z(t), \quad \text{where } Z(t)=\int_{0}^{t}F(s)ds.
\end{equation}
By substituting equation (\ref{Ch1}) into (\ref{BO1}), we obtain:
\begin{equation*}
\zeta_{t}+Z(t)\zeta_{x}+\zeta\zeta_{x}+\mathcal{H}[\zeta_{xx}]=0.
\end{equation*}
Next, we rewrite the equation in a moving frame:
\begin{equation*}\label{chv}
x \rightarrow x-V(t), \quad \text{where } V(t)=\int_{0}^{t}Z(s)ds,
\end{equation*}
which leads to the  BO equation with constant coefficients
\begin{equation}\label{BO1}
\zeta_{t}+\zeta\zeta_{x}+\mathcal{H}[\zeta_{xx}]=0.
\end{equation}
Equation (\ref{BO1}) admits algebraic soliton solutions \cite{Benjamin:1967, Choi:1996}
\begin{equation}\label{solitary}
\zeta(X) = \frac{A}{1+X^{2}/\Delta^{2}}, \quad \text{where } \Delta = \frac{4}{A}, \;\ X=x-ct \quad \text{with } c = -\frac{A}{4},
\end{equation}
where, $A$ represents the soliton amplitude, $\Delta$ the soliton width and $c$ its speed. Therefore, we have that
\begin{equation}\label{meanper000}
\eta(X,t)=Z(t)+\frac{A\Delta^{2}}{\Delta^{2}+(X-V(t))^{2}}
\end{equation}
are solutions of the random BO equation (\ref{meanper000}). This equation shows that a random external force mainly impacts the temporal variability of the pedestal supporting the soliton and the soliton speed, rather than causing changes in its amplitude. The pedestal, denoted as 
$Z(t)$, is spatially uniform and does not affect the soliton intrinsic properties.  

The mean field is obtained by statistically averaging the function $\eta$ in equation (\ref{meanper000}). It is convenient to normalize the interface $\eta$. This is done by rescaling the variables as follows
\begin{equation}\label{rescaling}
\eta \rightarrow \eta/A \;\ F \rightarrow  F/\Delta, \;\  X \rightarrow  X/\Delta.
\end{equation}
Replacing equation  (\ref{rescaling}) into equation (\ref{meanper000}) we obtain
\begin{equation}\label{meanper00}
\eta(X,t)=\frac{Z(t)}{4}+\frac{1}{1+(X-V(t))^{2}}.
\end{equation}

In numerous nonlinear dynamics problems, external forces often exhibit periodic or quasi-periodic behavior over time. Due to inherent fluctuations, the phase of these forces is frequently considered random. To account for this randomness, we introduce a zero-mean random state, represented as a Fourier series with $N$ harmonics. The external force is then described by
\begin{equation}\label{initial}
F(t) = \sum_{i=1}^{N}\sqrt{2S(\omega_i)\Delta\omega}\cos(\omega_{i}t+\varphi_{i}),
\end{equation}
 Here,  $S(\omega)$ represents the initial power spectrum, $\omega_i = i\Delta\omega$ with $\Delta\omega$ being the sampling frequency, and $\varphi_i$ denotes a random variable uniformly distributed in the interval $(0,2\pi)$. This assures that the external force is random.  We assume that the initial power spectrum has a Gaussian shape
\begin{equation}\label{Gaussian}
S(\omega) = Q\exp\Big(-\frac{(\omega-\omega_0)^2}{2K^2}\Big).
\end{equation}


The wave characteristics are determined by several parameters: the dimensionless peak number frequency ($\omega_0$), the spectral width denoted as $K$, and the relative energy indicated by the values of $Q$. The spectral band K is defined as $K=0.25$, the sampling frequency $\Delta\omega=0.05$  and the constant $Q$ is chosen such that  in all scenarios the total energy of the force has the same value, which is defined by the variance $\sigma^2 = 0.25$. The number of harmonics in the simulations is $N=256$, simulations were also performed with a higher number of harmonics, however, the results were qualitatively the same. 

Once $F(t)$ is chosen, the functions $Z(t)$ and  $V(t)$ can be obtained by integration. However, the constants of integration play a crucial role in affecting the soliton dynamics. Specifically, these constants determine different statistical properties of the soliton speed  $V(t)$ and the pedestal $Z(t)$. For instance, the phase $V(t)$ can be considered either as a stationary random process, including the initial time moment, or as a process that starts at $t>0$. Additionally, we also analyze $V(t)$ with growing dispersion, such as in a Brownian process. Moreover, our analysis begins with a case involving a purely periodic force, retaining only one harmonic in the equation (\ref{initial}).

\section{Results}
\subsection{$V(t)$ -- stationary random process}

\begin{figure}[h!]
	\centering	
	\includegraphics[scale =1]{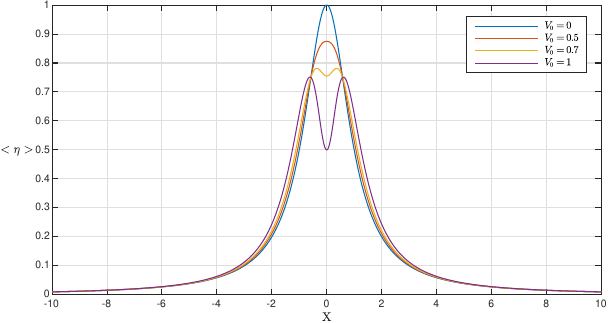}
	\caption{The asymptotics averaged soliton field given by equation (\ref{Taylor}) for different values $V_0$.}
	\label{split123}
\end{figure}

We initially assume the random force is represented by a one-mode Fourier component. This allows us to analyze the impact of a single frequency peak number on soliton dynamics.  The mean field is determined by averaging over the phase distribution function. Firstly, we consider the case when $V(t)$  represents a stationary random process for all times including $t = 0$.  In this case each initial soliton (\ref{meanper00}) has a random phase ($V(0)$). With this choice, we have that
\begin{equation}\label{onemode}
F(t) = F_{0}\cos(\omega_0 t+\varphi)
\end{equation}
where the frequency $\omega_0$ is fixed and $\varphi$ is a uniformly random variable distributed in the interval $(0,2\pi)$. The functions $Z(t)$ and $V(t)$ are also monochromatic
\begin{equation}
Z(t) = \frac{F_{0}}{\omega_0}\sin(\omega_0 t+\varphi), \mbox{ and } V(t) = -\frac{F_{0}}{\omega_0^2}\cos(\omega_0 t+\varphi).
\end{equation}
The soliton solution is written as 
\begin{equation}\label{meanper}
\eta(X,t)=\frac{F_{0}}{4\omega_0}\sin(\omega_0 t+\varphi)+\frac{1}{1+\Big(X+\frac{F_{0}}{\omega_{0}^2}\cos(\omega_0 t+\varphi)\Big)^{2}}.
\end{equation}
The averaged soliton field is given in terms of the integral
\begin{equation}\label{meanper123}
<\eta(X,t)>=\frac{1}{2\pi}\int_{0}^{2\pi}\Bigg[\frac{1}{1+\Big(X+\frac{F_{0}}{\omega_0^2}\cos(\omega_0 t+\varphi)\Big)^{2}}\Bigg]d\varphi.
\end{equation}
This integral does not have a simple closed form. However, by defining $V_0=F_0/\omega_0^2$ and assuming that $V_0$ is small, we can compute the integral (\ref{meanper}) asymptotically. Using the Taylor series we have that
\begin{equation}\label{exp}
\frac{1}{\Big(1+(X+V_0\cos(\omega_0 t+\varphi))^{2} \Big)}=\frac{1}{1+X^2}-V_0\frac{2X\cos(\omega_0 t+\varphi)}{(1+X^2)^2}+V_0^{2}\frac{(3X^2-1)\cos^2(\omega_0 t+\varphi)}{(1+X^2)^3}+\mathcal{O}(V_0^3).
\end{equation}
Consequenltly, substituting equation (\ref{exp}) into (\ref{meanper123}) we obtain
\begin{equation}\label{Taylor}
<\eta(X,t)>=\frac{1}{1+X^2}+V_0^{2}\frac{3X^2-1}{2(1+X^2)^3}.
\end{equation}
Notice that at $X=0$ we have that
\begin{equation*}
<\eta(0,t)>= 1-\frac{V_0^2}{2}.
\end{equation*}
This means that the force frequency dampens the soliton averaged amplitude; however, it remains constant over time. In particular, as $V_0\rightarrow 0^{+}$, the soliton averaged amplitude resembles an algebraic soliton solution of the BO equation. Another asymptotic can be done in the vicinity of the soliton crest for any values of $V_0$. The Taylor series expansion determine the soliton averaged behavior near $X=0$. The Taylor series yields
\begin{equation}
<\eta(X,t)>= 1-\frac{V_0^2}{2} +(3V_0^2-1)X^2.
\end{equation}
Thus, if $V_0>1/\sqrt{3}$ the soliton field is split into two, otherwise a single peak arises. This is  illustrated in Figure \ref{split123}  for different values of $V_0$.

\begin{figure}[h!]
	\centering	
	\includegraphics[scale =1]{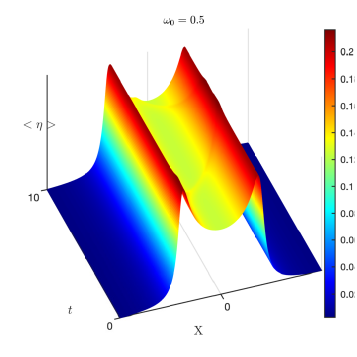}
	\includegraphics[scale =1]{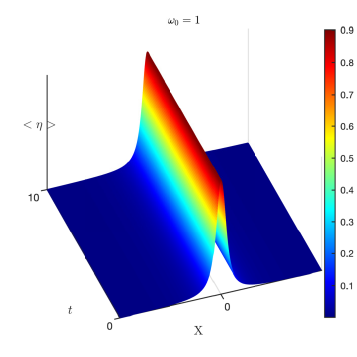}
	\caption{The averaged soliton field given by equation (\ref{meanper}) frequency peak number ($\omega_0$) with $F_0=0.5$.}
	\label{periodic2}
\end{figure}

Now, we compute  the integral  (\ref{meanper123}) numerically using the trapezoidal rule to obtain the soliton averaged field.
Figure \ref{periodic2} presents the spatio-temporal diagrams of the averaged soliton fields at various frequencies. As observed, the averaged soliton consists of two steady pulses with damped amplitude. These results agree with the asymptotic analysis presented above. For instance, in Figure \ref{periodic2} (left), where  $V_0=1$ the soliton splits into two. In contrast, Figure \ref{periodic2} (right) shows a single soliton for $V_0=0.5$. The soliton averaged splitting is a phenomenon is similar to the one described in \cite{Pelinovsky:2006}. 

After discussing the case of a single Fourier harmonics, we assume that the random force is given by the Fourier series described in equation (\ref{initial}). The functions $Z(t)$ and $V(t)$ have similar form

\begin{equation}
Z(t) =  \sum_{i=1}^{N}\frac{\sqrt{2S(\omega_i)\Delta\omega}}{\omega_i}\sin(\omega_{i}t+\varphi_{i}) \mbox{ and }
V(t) =  -\sum_{i=1}^{N}\frac{\sqrt{2S(\omega_i)\Delta\omega}}{\omega_i^2}\cos(\omega_{i}t+\varphi_{i}).
\end{equation}
In this case the averaged soliton field can only be obtained numerically. However, it is not necessary to solve the BO equation (\ref{BO1}). This is done by averaging equation (\ref{meanper000}) for a large number of realizations. 
\begin{figure}[h!]
	\centering	
	\includegraphics[scale =1]{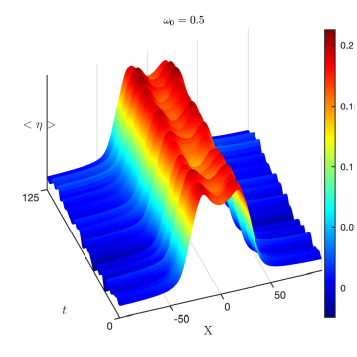}
	\includegraphics[scale =1]{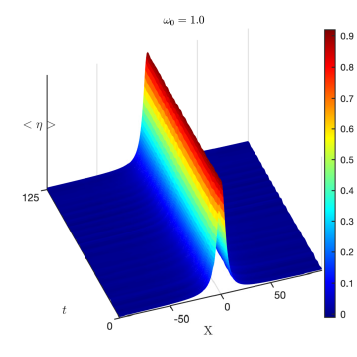}
	\includegraphics[scale =1]{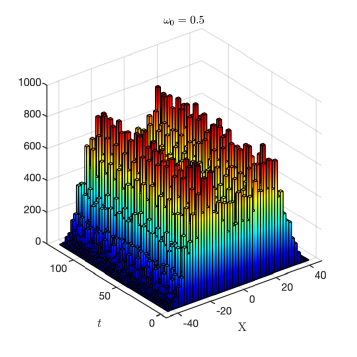}
	\includegraphics[scale =1]{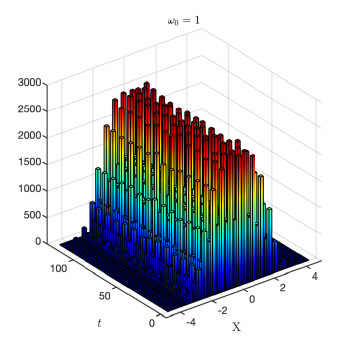}
	\caption{Top: the averaged soliton field over 10,000 realizations for different frequency peak number ($\omega_0$) and their respective histogram (bottom).}
	\label{periodic1}
\end{figure}

Figure \ref{periodic1} (top) illustrates the averaged soliton field over   $10,000$ realizations for different frequency peak numbers.  In order to assure that the results are statistical significant we compare the results with ensemble of $5000$ and statistically, the results are the same. At low frequencies ($\omega_0 < 1$), the soliton averaged field has the tendency  to split into two, whereas at higher frequencies, the amplitude fluctuations are minimal. Unlike the case where $V$ follows a normal distribution or a uniform distribution within a symmetric interval, leading to both damping and spreading of the soliton averaged field see for instance \cite{Zahibo:2009}. Results are in agreement with the asymptotic theory. Another justification for the soliton split at  $t=0$ is the randomness of $V(0)$, therefore the soliton crest might be shifted from the origin at $t=0$. Notably, the distribution of probability of $\eta$ exhibits bimodal behavior at lower frequencies and unimodal behavior at higher frequencies. In Figure \ref{periodic1} (bottom), the histogram shows the number of soliton crests at position $X$ and time $t$ for 10,000 realizations. This histogram confirms that the soliton has "two preferred" trajectories. At higher frequencies, soliton crest trajectory remains almost constant and the soliton averaged field resembles the soliton solution of the BO equation, see Figure \ref{periodic1} (right).

\subsection{$V(0) = 0$ -- the case of switch on the random $V(t)$}
In this section, we briefly discuss the case of a single Fourier mode where the initial soliton random phase $(V(t))$ is switched on at $t=0$. In other words, the initial soliton crest is always located at $X=0$ at time $t=0$. This condition is achieved by selecting the external force as described earlier in equation (\ref{onemode}). In this case the random phase is
\begin{equation}
V(t) = -\frac{F_{0}}{\omega_0^2}\cos(\omega_0 t+\varphi)+\frac{F_{0}}{\omega_0^2}\cos(\varphi).
\end{equation}
Therefore, we can compute the soliton averaged field through the integral
\begin{equation}\label{meanper0012}
<\eta(X,t)>=\frac{1}{2\pi}\int_{0}^{2\pi}\Bigg[\frac{1}{1+\Big(X+\frac{F_{0}}{\omega_0^2}\cos(\omega_0 t+\varphi)-\frac{F_{0}}{\omega_0^2}\cos(\varphi)\Big)^{2}}\Bigg]d\varphi.
\end{equation}
\begin{figure}[h!]
	\centering	
	\includegraphics[scale =1]{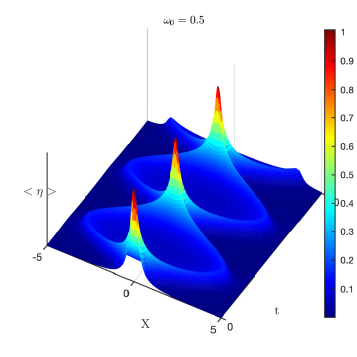}
	\includegraphics[scale =1]{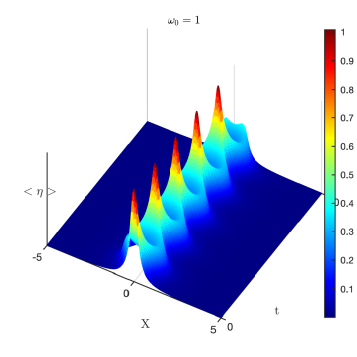}
	\caption{The averaged soliton field given by equation (\ref{meanper0012}) frequency peak number ($\omega_0$) with $F_0=0.5$.}
	\label{Det}
\end{figure}

As in the previous section, defining $V_0=F_0/\omega_0^2$  and assuming that $V_0\ll 1$ and making use of the Taylor series yields
\begin{equation} \label{exp1}
\begin{split}
\frac{1}{\Big(1+(X+V_0(\cos(\omega_0 t+\varphi)-\cos(\varphi)))^{2} \Big)}= & \frac{1}{1+X^2}-V_0\frac{2X(\cos(\omega_0 t+\varphi)-\cos(\varphi))}{(1+X^2)^2} + \\
&+ V_0^{2}\frac{(3X^2-1)(\cos(\omega_0 t+\varphi)-\cos(\varphi))^2}{(1+X^2)^3}+\mathcal{O}(V_0^3).
\end{split}
\end{equation}
Thus, substituting (\ref{exp1}) into (\ref{meanper0012}) we obtain
\begin{equation}\label{Taylor}
<\eta(X,t)>=\frac{1}{1+X^2}+V_0^{2}\frac{3X^2-1}{(1+X^2)^3}(1-\cos(\omega_0t)).
\end{equation}
Similarly, using the Taylor series near $X=0$ for arbitrary $V_0$ yields
\begin{equation}\label{Taylor2}
<\eta(X,t)>= 1-V_0^2(1-\cos(\omega_0t))+(6V_0^2(1-\cos(\omega_0t))-1)X^2
\end{equation}
Therefore, the soliton averaged field split into two and join together periodically as long as
\begin{equation}
6V_0^2(1-\cos(\omega_0t))-1=0.
\end{equation}
Thus, the periodic motion changes at 
\begin{equation}\label{tc}
t=\cos^{-1}(1-1/6V_0^2)/\omega_0+n\pi, \mbox{ where $n$ is an integer.} 
\end{equation}

The predictions are confirmed numerically computing (\ref{meanper0012}). Figure \ref{Det} displays the soliton averaged field for different frequency peak-numbers. The periodic motion predicted by the asymptotic theory is confirmed.

Notice yet that at $X=0$ equation (\ref{Taylor2}) yields 
 \begin{equation}\label{Taylor2}
<\eta(0,t)>= 1-V_0(1-\cos(\omega_0t)),
\end{equation}
Thus, at $t=2n\pi/\omega_0$, for $n$ integer, the averaged soliton field attains its maximum, while at $X=0$  the minimum value occurs at $t=(2n-1)2\pi/\omega_0$, for $n$ integer.

\subsection{$V(t)$ -- random process with growing dispersion}
In this section, we no longer set the integral constants to be zero and in this case $V(t)$ has the secular term as it is typical for the Brownian processes. For the one-mode Fourier problem we have equation (\ref{onemode}) and the functions $Z(t)$ and $V(t)$ are \begin{equation}
Z(t) = \frac{F_{0}}{\omega_0}\sin(\omega_0 t+\varphi)-\frac{F_{0}}{\omega_0}\sin(\varphi), \mbox{ and }
\end{equation}
\begin{equation}
V(t) = -\frac{F_{0}}{\omega_0^2}\cos(\omega_0 t+\varphi)+\frac{F_{0}}{\omega_0^2}\cos(\varphi)-t\frac{F_{0}}{\omega_0}\sin(\varphi).
\end{equation}
Here,  $V(0) = 0$ and therefore all initial solitons have the same phase and $V(t)$ has growing dispersion as in the Brownian processes. The soliton averaged field is computed through the formula
\begin{equation}\label{meanper1214}
<\eta(X,t)>=\frac{1}{2\pi}\int_{0}^{2\pi}\Bigg[\frac{1}{1+\Big(X+\frac{F_{0}}{\omega_0^2}\cos(\omega_0 t+\varphi)-\frac{F_{0}}{\omega_0^2}\cos(\varphi)+t\frac{F_{0}}{\omega_0}\sin(\varphi)\Big)^{2}}\Bigg]d\varphi.
\end{equation}

Asymptotics can also be performed in a similar fashion as in the previous section. We consider $V_0=F_0/\omega_0$ and assume that $\omega_0 t\gg 1$. Under these assumptions we have that $tV_0=tF_{0}/\omega_0\gg F_{0}/\omega_0^2$, consequently, we can approximate the integral (\ref{meanper1214}) as
\begin{equation}\label{meanper}
<\eta(X,t)>=\frac{1}{2\pi}\int_{0}^{2\pi}\Bigg[\frac{1}{1+\Big(X+V_0 t\sin(\varphi)\Big)^{2}}\Bigg]d\varphi.
\end{equation}
This integral can be computed in exactly form at $X=0$. Thus,
\begin{equation}\label{meanperX0}
<\eta(0,t)>=\frac{1}{2\pi}\int_{0}^{2\pi}\Bigg[\frac{1}{1+\Big(V_0 t\sin(\varphi)\Big)^{2}}\Bigg]d\varphi=\frac{1}{\sqrt{1+V_0t}}
\end{equation}
This shows that the soliton amplitude decays over time. 

Another regime can be considered to provide more details of the soliton averaged field near 
$X=0$. Let us now assume that  $V_0=F_0/\omega_0^2$ is small and that $\omega_0 t$ is order $\mathcal{O}(1)$. In this case,  equation (\ref{meanper1214}) takes the form
\begin{equation}\label{meanper13}
<\eta(X,t)>=  \frac{1}{2\pi}\int_{0}^{2\pi}\Bigg[\frac{1}{1+\Big(X+V_0\cos(\omega_0 t+\varphi)-V_0\cos(\varphi)+V_0 t\omega_0\sin(\varphi)\Big)^{2}}\Bigg]d\varphi,
\end{equation}
and we can consider a Taylor expansion near $V_0=0$ this yields
\begin{equation} \label{exp2}
\begin{split}
\frac{1}{1+(X+V_0\cos(\omega_0 t+\varphi)-V_0\cos(\varphi)+V_0\omega_0t\sin(\varphi))^{2} } &=  \frac{1}{1+X^2}-V_0\frac{2X(\cos(\omega_0 t+\varphi)-\cos(\varphi)+\omega_0t\sin(\varphi))}{(1+X^2)^2} \\
&+ V_0^{2}\frac{(3X^2-1)(\cos(\omega_0 t+\varphi)-\cos(\varphi)+\omega_0t\sin(\varphi))^2}{(1+X^2)^3}.
\end{split}
\end{equation}
Substituting (\ref{exp2}) into (\ref{meanper1214})  we have 
\begin{equation}\label{Taylor}
<\eta(X,t)>=\frac{1}{1+X^2}+V_0^{2}\frac{3X^2-1}{2(1+X^2)^3}\Big(\omega_0^2t^2+2-2\omega_0t\sin(\omega_0t)-2\cos(\omega_0t)\Big).
\end{equation}
In particular, near $X=0$, 
\begin{equation}\label{Taylor3}
<\eta(X,t)>= 1-\frac{V_0^2}{2}\Big(\omega_0^2t^2+2-2\omega_0t\sin(\omega_0t)-2\cos(\omega_0t)\Big)+\Big[3V_0^{2}\Big(\omega_0^2t^2+2-2\omega_0t\sin(\omega_0t)-2\cos(\omega_0t)\Big)-1\Big]X^2
\end{equation}
Let $W = \omega_0 t$ and define
\begin{equation}
G(W)=W^2+2-2W\sin(W)-2\cos(W),
\end{equation} 
Thus, $G$ is a non-decreasing function for $W \ge 0$, with its range being $[0, +\infty)$. Mathematically, this implies that the coefficient of $X^2$ in equation (\ref{Taylor3}) can change sign from negative to positive only once. Physically, this means that once the soliton splits into two, they never rejoin, which is different from the results obtained in the previous section.


\begin{figure}[h!]
	\centering	
	\includegraphics[scale =0.95]{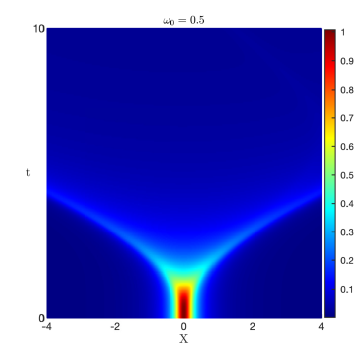}
	\includegraphics[scale =0.95]{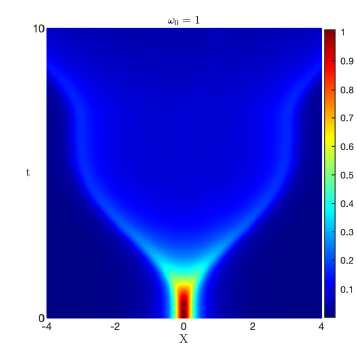}
	\includegraphics[scale =0.95]{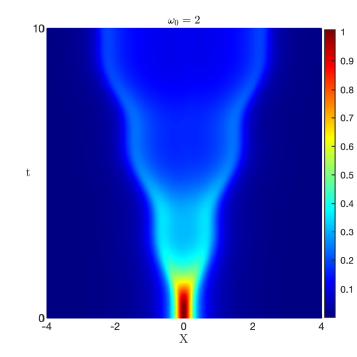}
	\includegraphics[scale =0.95]{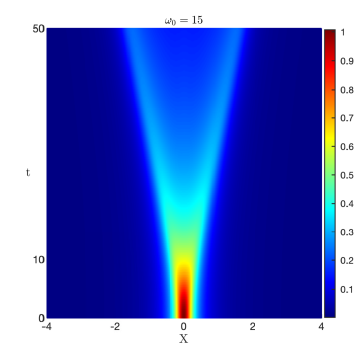}
	\caption{The averaged soliton field given by equation (\ref{meanper}) frequency peak number ($\omega_0$) with $F_0=0.5$.}
	\label{periodic4}
\end{figure}
For any time, the integral given by equation (\ref{meanper1214}) is computed numerically through the trapezoidal rule. Figure \ref{periodic4} displays spatio-temporal diagrams of the averaged soliton fields at different frequencies. The diagrams illustrate that the averaged soliton splits into two pulses, each propagating at the same speed but in opposite directions which agree well with the asymptotic predictions. Besides,  the averaged soliton splitting is qualitatively similar to the phenomenon reported in \cite{Pelinovsky:2006}. However, a notable distinction in this context is that, although the two solitons move at identical speeds in opposite directions, their velocities are not necessarily constant, unlike the constant velocities observed in the KdV framework discussed in \cite{Pelinovsky:2006}.
\begin{figure}[h!]
	\centering	
	\includegraphics[scale =1]{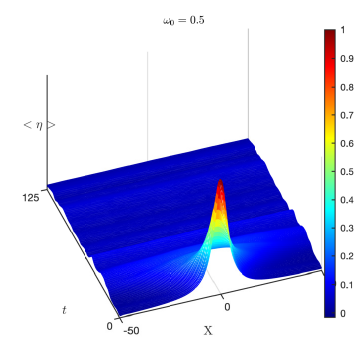}
	\includegraphics[scale =1]{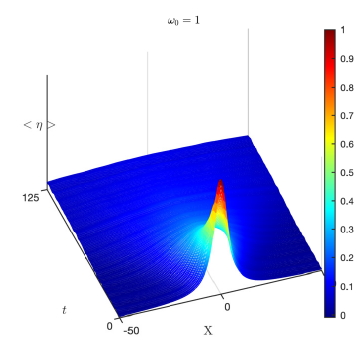}
	\includegraphics[scale =1]{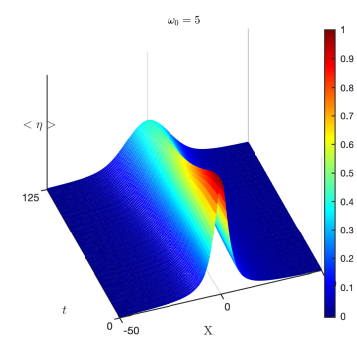}	
	\includegraphics[scale =1]{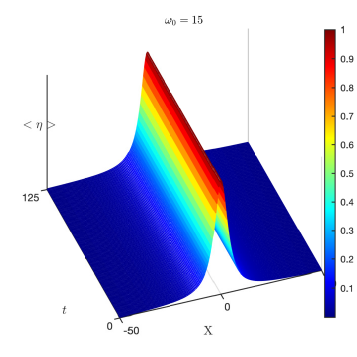}
	\caption{The averaged soliton field over 10,000 realizations for different frequency peak number ($\omega_0$).}
	\label{periodic3}
\end{figure}

Now, assuming  $F(t)$ to be given by Fourier series  as in equation (\ref{initial}) the function $V(t)$ can be obtained in the form
\begin{equation}
V(t) =  -\sum_{i=1}^{N}\frac{\sqrt{2S(\omega_i)\Delta\omega}}{\omega_i^2}\cos(\omega_{i}t+\varphi_{i})- t\sum_{i=1}^{N}\frac{\sqrt{2S(\omega_i)\Delta\omega}}{\omega_{i}}\sin(\varphi_{i}) + \sum_{i=1}^{N}\frac{\sqrt{2S(\omega_i)\Delta\omega}}{\omega_i^2}\cos(\varphi_{i})
\end{equation}
With these choices $V(t)$ is the Brownian process and $V(0) = 0$ at $t = 0.$

\begin{figure}[h!]
	\centering	
	\includegraphics[scale =1]{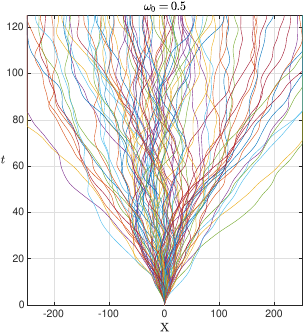}
	\includegraphics[scale =1]{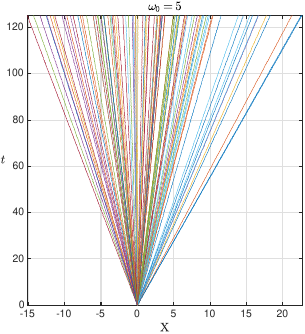}
	\caption{The  soliton crest trajectories for different 100 realizations and different frequency peak numbers ($\omega_0$).}
	\label{trajectories}
\end{figure}

\begin{figure}[h!]
	\centering	
	\includegraphics[scale =1]{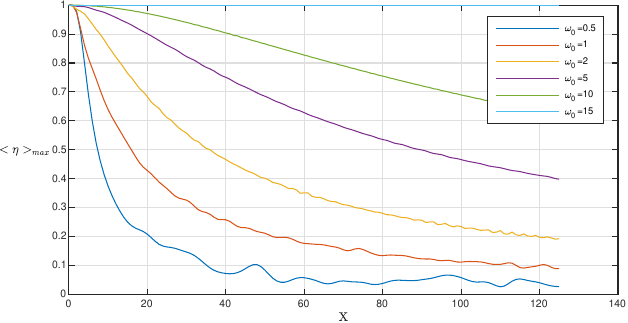}
	\caption{ The averaged over 10,000 realizations of the variation of amplitude of mean soliton field $<\eta(0,t)>$ for different frequencies.}
	\label{etamax1}
\end{figure}

The averaged soliton field is obtained numerically 
by averaging equation (\ref{meanper00}) over $10,000$ realizations. Figure \ref{periodic3} illustrates the averaged soliton field over $10,000$ realizations for various frequency peak numbers. At low frequencies ($\omega_0 < 1$), there is a significant decay in the soliton averaged amplitude, whereas at higher frequencies, the amplitude fluctuations are minimal.  This mainly occurs due to the crest trajectories, as it can be seen in Figure \ref{trajectories}. At low frequencies, the soliton tend to spread more than at high frequencies, which affects the expected value of the soliton field. This agrees partially with the asymptotic theory. Here though, it is not clear whether or not the soliton field splits into two at large times. From our computations, at large times a soliton field does not have well defined peaks. Moreover, the decay rate is highly dependent on the frequency peak number and faster than the asymptotic predictions, see Figure \ref{etamax1}. 

\section{Conclusion}
In this work, we have investigated algebraic soliton interactions with a time-dependent periodic random external force using the Benjamin-Ono equation. The stochastic solution of the forced BO equation is found in explicit form and then averaged on random phase.  We computed the expected value of the soliton wave field asymptotically and compared it with numerical results, showing strong agreement. In Sect. 3.1, we considered the soliton phase as a random process and demonstrated that the averaged soliton field can resemble a BO-soliton if the frequency peak-number is high or consists of two steady solitons. In Sect. 3.2, we examined the case where the initial soliton phase is deterministic, and randomness is introduced at $t = 0^+$. In this scenario, an interesting phenomenon occurs: the soliton averaged field splits into two pulses, then rejoins to almost return to its initial form. This motion is repeated periodically over time.  Lastly, in Sect. 3.3, we assumed the phase to be a process with growing dispersion (Brownian process). We showed that the soliton averaged field amplitude dampens over time, leading to two possible scenarios: (i) the soliton averaged field always resembles a BO-soliton with its amplitude decaying over time, or (ii) the soliton averaged field splits into two solitons that travel in opposite directions with variable speeds.

%

\section{Acknowledgements}
E.P. and E.D. were supported by Laboratory of Nonlinear Hydrophysics and Natural Disasters of the V.I. Il’ichev Pacific Oceanological Institute, grant from the Ministry of Science and Higher Education of the Russian Federation, agreement number 075-15-2022- 1127 from 01.07.2022.

	\section*{Declarations}
	
	\subsection*{Conflict of interest}
	The authors state that there is no conflict of interest. 
	\subsection*{Data availability}
	
	Data sharing is not applicable to this article as all parameters used in the numerical experiments are informed in this paper.

\end{document}